\documentclass[aps,preprintnumbers, twocolumn,reprint,superscriptaddress,nofootinbib]{revtex4}
\usepackage{amsmath}
\usepackage{graphicx}
\usepackage{amssymb}

\newcommand{\be}{\begin{equation}}
\newcommand{\ee}{\end{equation}}
\newcommand{\bea}{\begin{eqnarray}}
\newcommand{\eea}{\end{eqnarray}}
\newcommand{\met}{\not{\!\!{\rm E}}_{T}}

\begin{document}

\preprint{ANL-HEP-PR-11-44, MSUHEP-110725, NSF-KITP-11-131}

\title{Calculation of Associated Production of a Top Quark and a $W'$ at the LHC}

\author{Edmond L. Berger}
\affiliation{High Energy Physics Division, Argonne National Laboratory, 
Argonne, IL 60439, U.S.A}

\author{Qing-Hong Cao}
\affiliation{High Energy Physics Division, Argonne National Laboratory, Argonne, IL 60439, U.S.A}
\affiliation{Enrico Fermi Institute, University of Chicago, Chicago, Illinois 60637, U.S.A.}
\affiliation{Department of Physics and State Key Laboratory of Nuclear Physics
and Technology, Peking University, Beijing 100871, China}

\author{Jiang-Hao Yu}
\affiliation{Department of Physics and Astronomy, Michigan State University, 
East Lansing, MI 48823, U.S.A}

\author{C.-P. Yuan}
\affiliation{Department of Physics and Astronomy, Michigan State University, 
East Lansing, MI 48823, U.S.A}
\affiliation{Center for High Energy Physics, Peking University, Beijing 100871, China}

\begin{abstract}
	
We investigate collider signatures of a top-philic $W^\prime$ model, 
in which the $W^\prime$ boson couples only to the third-generation quarks of 
the standard model. 
The main discovery channel for this $W^\prime$ is through associated 
production of the $W^\prime$ and top quark, yielding a top-quark pair 
plus an extra bottom quark jet as a signal.   We do a full simulation 
of the signal and relevant backgrounds.   We develop a method of analysis 
that allows us to conclude that discovery of the $W^\prime$ is promising 
at the LHC despite large standard model backgrounds.   Bottom quark 
tagging of the extra jet is key to suppressing the backgrounds.   
\end{abstract}
 
\maketitle

\section{Introduction}

Extra charged gauge bosons ($W^\prime$'s) are present in models of 
new physics (NP) beyond the standard model (SM) and often assumed 
to be produced as direct $s$-channel resonances in hadron collisions.   
Searches have been carried out at the Tevatron and at the Large Hadron 
Collider(LHC) for $s$-channel $W^\prime$'s in lepton decay 
modes~\cite{Khachatryan:2010fa,Aad:2011fe,Aaltonen:2010jj},
in single top-quark channels~\cite{Aaltonen:2009qu}, and in diboson 
decays~\cite{Abazov:2010dj}.   

In this paper we study a different  $W^\prime$ model, named a ``top-philic'' model, 
in which the $W^\prime$ couples only to third-generation quarks (top and bottom 
quarks) and is  produced only in association with a top quark.   It decays only into 
a top quark and a bottom quark pair.  The collider signature of the events is a 
$t\bar{t}$ pair plus one $b$-jet.   We explore the discovery 
potential of  the top-philic $W^\prime$ boson at the LHC at a center of mass 
energy of 14~TeV with an integrated luminosity of $100~{\rm fb}^{-1}$.  We 
compute the inclusive cross section, simulate the signal and backgrounds, 
and investigate a set of optimal cuts.   
Our study shows that the prospects are promising to discover the 
top-philic $W^\prime$ in $tW^\prime$ associated production despite the 
presence of SM backgrounds that exceed our signal by three or four orders of 
magnitude.   The key is to identify as a $b$ jet the extra jet produced in association with 
$t\bar{t}$.   A 1~TeV $W^\prime$ with the same coupling strength as the 
SM $W$-$t$-$b$ interaction could be discovered with a $5$ standard deviations 
statistical significance at the LHC at 14~TeV.  

Motivated by the observation of large parity violation in top quark 
pair-production at the Tevatron~\cite{Aaltonen:2011kc}, several authors
have recently proposed a $W^\prime$ boson with a flavor changing $d$-$t$-$W^\prime$ 
interaction~\cite{Barger:2010mw,Cao:2010zb,Barger:2011ih}. 
This $W^\prime$ boson is also produced in association with a top quark, but it 
differs from the top-philic $W^\prime$ we discuss in that it decays into a top quark 
and a non-$b$ quark, yielding a final state of $t\bar{t}$ plus a non-$b$ 
jet~\cite{Gresham:2011dg}.   This final state suffers from a huge $t\bar{t}j$ background 
that cannot be mitigated by $b$-tagging on the jet produced in association with the  
top quark pair.   As a result, a large coupling strength would be needed for discovery  
of the flavor changing $W^\prime$ at the LHC.
 
\section{The model}

A top-philic $W^\prime$ can arise from a new non-abelian gauge symmetry which breaks
generation universality~\cite{Malkawi:1996fs,Li:1981nk,He:2002ha}. 
A summary may be found in Ref.~\cite{Hsieh:2010zr}. 
In this study we adopt an effective Lagrangian approach rather than focusing 
on specific NP models. 
The effective, renormalizable interaction of the $W'$ to the SM third generation fermions 
is
\be
{\cal L} = i \frac{g_2}{\sqrt{2}}~\bar{t} \gamma^\mu (f_{L} P_L +  f_{R} P_R) b 
~W^{\prime +}_\mu + {\rm h.c.}~~, 
\ee
where $g_2=e/\sin\theta_W$ is the weak coupling, while $P_{L/R}$ are the 
usual chirality projection operators.
For simplicity, we consider only the case with a purely left-handed 
current ($f_{L} = 1$, and $f_{R} = 0$),  
but our study can be extended easily to other cases.   
The triple gauge interaction of the $W'$ and SM gauge bosons is not included 
because such a non-abelian interaction is suppressed for large $W^\prime$ mass ($m_{W'}$) 
by $W$-$W^\prime$ mixing effects, which are of order ${\cal{O}}({m_W^2}/{m_{W^\prime}^2})$.

The $W^\prime$ decays entirely to a top quark and bottom quark pair with decay width  
\be
\Gamma_{W^\prime \to t \bar{b}}\simeq\frac{3g_2^2 m_{W^\prime }}{48\pi}\left(f_{L}^2+f_{R}^2\right).
\ee
For $f^2_{L}+f^2_{R}=1$, $\Gamma_{W^\prime}\sim m_{W^\prime}/100$, indicating that the 
$W'$ is quite narrow.  

\begin{figure}
	\includegraphics[scale=0.4]{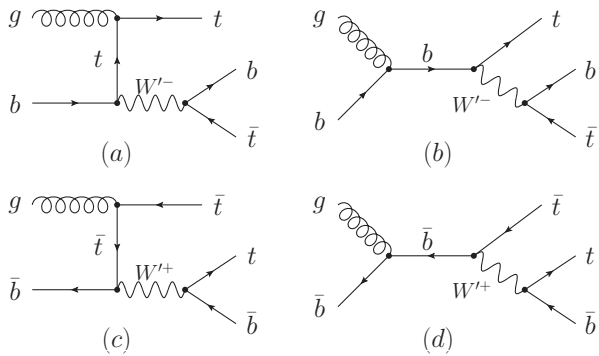}
	\caption{Feynman diagrams for the associated production of a $W^\prime$ and a top 
	quark: (a, b) $W^{\prime-}t$, and (c, d) $W^{\prime+}\bar{t}$.}
	\label{fig:feyndiag}
\end{figure}

The top-philic $W^\prime$ boson is produced predominately through
a gluon-bottom-quark fusion process, as depicted in Fig.~\ref{fig:feyndiag}. 
The $W^\prime$ decays into a top and bottom quark pair, and the overall 
final state is then $t\bar{t}b$.  Since this final state has not been used to search for a 
$W^\prime$ at the Tevatron and LHC, none of the current collider limits 
constrain our top-philic $W^\prime$ model.   It is possible the top-philic 
$W^\prime$ boson could be as light as a few hundred GeV. 

The recent CDF measurement of the ratio of the cross sections for 
$t\bar{t}+0$~jets to $t\bar{t}+n$ jets is consistent with the SM 
expectation~\cite{Aaltonen:2009iz}.  Top-philic 
$W^\prime$ production will contribute to the $t\bar{t}+n$~jets rate.  However, 
our numerical calculation shows that $tW^\prime$ production is too small to be 
of concern, e.g. $\sigma(tW^{\prime-}+\bar{t}W^{\prime+}) \sim 3~{\rm fb}$ for 
$m_{W^\prime}=200~{\rm GeV}$, $f_L=1$, and $f_R=0$.  Moreover, the 
cross section drops rapidly with $m_{W^\prime}$.   We conclude that the 
top-philic $W^\prime$ model is consistent with $t\bar{t}$ current measurements at 
the Tevatron.

\begin{figure}
\includegraphics[scale=0.35]{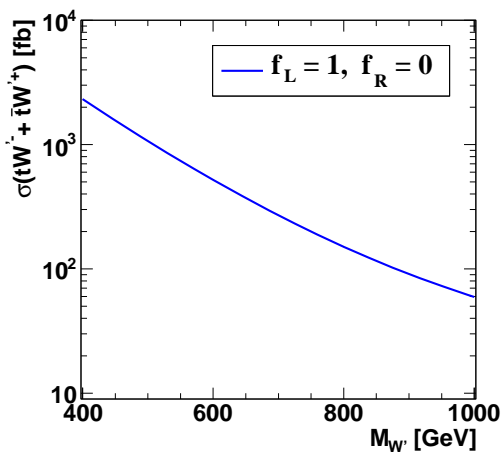}
\caption{The inclusive cross section of the $tW^{\prime-}$ and $\bar{t}W^{\prime +}$ 
associated production at a 14~TeV LHC with $f_{L}=1$ and $f_{R}=0$.}
\label{fig:crosec}
\end{figure}

In Fig.~\ref{fig:crosec}, we display the leading-order inclusive cross section for  
$tW^{\prime-}$ and $\bar{t}W^{\prime +}$ production, 
$\sigma(tW^{\prime-}+\bar{t}W^{\prime+})$, as a function of the $W^\prime$ 
mass ($m_{W^\prime}$) at 14~TeV with
a purely left-handed $W^\prime$-$t$-$b$ coupling, i.e. $f_{L}=1$ and $f_{R}=0$.
Note that $\sigma(\bar{t}W^{\prime +})=\sigma(tW^{\prime -})$
owing to equality of the parton distribution functions for initial state $b$- and 
$\bar{b}$-quarks.  
The CTEQ6L parton distribution functions are used in our calculation 
with the renormalization and factorization scales chosen as $m_{W^\prime}$. 
The production cross section is at the picobarn level for a $W^\prime$ with a 
few hundred GeV mass and at the femtobarn level for a multiple 
TeV-scale $W^\prime$.

\section{Collider phenomenology}

Our signal consists of both $tW^{\prime-}$ and $\bar{t}W^{\prime+}$ production channels 
because the two channels give rise to a same $t\bar{t}$ plus one $b$-jet collider signature 
 after the $W^\prime$ decay.  The $b$-jet could originate from a $b$ or a $\bar{b}$ as 
 one cannot now distinguish $b$- and $\bar{b}$-jets experimentally.  
At the event reconstruction level the $b$-jet together with one top quark would yield
a heavy $W^\prime$ resonance. 
The main SM background is from production of a $t\bar{t}$ pair plus one $b$ jet.  We also take 
into account the possibility that a light quark jet fakes a $b$ jet. 
The signal and background events are generated with MadGraph5/MadEvent~\cite{Alwall:2011uj}. 

In order to trigger on the signal event, we demand a leptonic decay of the top quark 
$t\to b\ell^+\nu_{\ell}$ and hadronic decay of the antitop $\bar{t}\to \bar{b}jj$.
The signal processes are  
\begin{eqnarray}
pp&\to& tW^{\prime-} \to  t\bar{t}b \to bW^{+}\bar{b} W^{-}b \to 
bb\bar{b}\ell^+ jj \nu ,\nonumber \\
pp&\to& \bar{t}W^{\prime+} \to  t\bar{t}\bar{b} \to bW^{+}\bar{b}  W^{-}\bar{b}
\to b\bar{b}\bar{b}\ell^+ jj \nu.
\label{eq:signal}
\end{eqnarray}
The topology of our signal is characterized by one isolated positive charged lepton, 
five high energy jets, and a large missing transverse momentum ($\met$) from the missing 
neutrino.  Both electrons and muons are used in our analysis.  

\begin{table}
\caption{The numbers of signal and background events at 14~TeV 
with an integrated luminosity of 100 fb$^{-1}$ before and after cuts, with $f_L=1$, 
for seven values of $m_{W^\prime}$~(GeV).   The top quark decay branching ratio 
2/27 is included in the ``no cut" column, and the $b$ tagging efficiency is included 
in the fifth column.  
The cut acceptances $\epsilon_{\rm cut}$ are also listed.
\label{tab:xsec}}
\begin{tabular}{ccccccc}
\hline\hline
$m_{W^\prime}$ & No cut  &  {\it basic} & {\it optimal} & {\it b-tagging} & $\Delta M$ cut &  $\epsilon_{\rm cut}$  \tabularnewline 
\hline\hline
400          &  32920  & 6929    & 5240   & 3018 & 2166 &  6.6 \% \tabularnewline
 $t\bar{t}b$ &  1.9$\times 10^{5}$    & 23849            & 2712   & 1537 & 297 &  0.15 \% \tabularnewline
 $t\bar{t}j$ & 3.13$\times 10^{7}$  & 3$\times 10^{6}$   & 306062 & 6984 & 967 &  3.1$\times 10^{-3}$\%   \tabularnewline
\hline
500          &  15115  & 3324    & 2621  & 1513 & 1120 &  7.4 \% \tabularnewline
 $t\bar{t}b$ &  1.9$\times 10^{5}$    & 23849             & 2709   & 1529 & 449 &  0.23 \% \tabularnewline
 $t\bar{t}j$ & 3.13$\times 10^{7}$  & 3$\times 10^{6}$    & 306057 & 6895 & 577 &  1.8 $\times 10^{-3}$\%   \tabularnewline
\hline
600          &  7361  & 1666     & 1300   & 754 & 565 &  7.7 \% \tabularnewline
 $t\bar{t}b$ &  1.9$\times 10^{5}$    & 23849             & 2524  & 1429 & 437 &  0.23 \% \tabularnewline
 $t\bar{t}j$ & 3.13$\times 10^{7}$  & 3$\times 10^{6}$    & 288098 & 6214 & 385 &  1.2 $\times 10^{-3}$\%   \tabularnewline
\hline
700          &  3843   & 874    & 638   & 369  & 282 &  7.4 \% \tabularnewline
 $t\bar{t}b$ &  1.9$\times 10^{5}$    & 23849            & 1781   & 1026 & 303 &  0.16 \% \tabularnewline
 $t\bar{t}j$ & 3.13$\times 10^{7}$  & 3$\times 10^{6}$   & 212153 & 4441 & 304 &  9.7 $\times 10^{-4}$\%   \tabularnewline
\hline
800          &  2110   & 490      & 405    & 197  & 154  &  7.3 \% \tabularnewline
 $t\bar{t}b$ &  1.9$\times 10^{5}$    & 23849            & 1060  & 620  & 189  &  0.10 \% \tabularnewline
 $t\bar{t}j$ & 3.13$\times 10^{7}$  & 3$\times 10^{6}$   & 130122 & 2346  & 214  &  6.8 $\times 10^{-4}$\%   \tabularnewline
\hline
900          &  1215   & 290      & 187   & 107 & 85  &  6.9 \% \tabularnewline
 $t\bar{t}b$ &  1.9$\times 10^{5}$    & 23849            & 594    & 353  & 110  &  0.058 \% \tabularnewline
 $t\bar{t}j$ & 3.13$\times 10^{7}$  & 3$\times 10^{6}$   & 74342  & 1052  & 62  &  2.0 $\times 10^{-4}$\%   \tabularnewline
\hline
1000          &  720    & 172     & 106    & 62   & 50  &  7.0 \% \tabularnewline
 $t\bar{t}b$ &  1.9$\times 10^{5}$    & 23849           & 337    & 199  & 64  &  0.034 \% \tabularnewline
 $t\bar{t}j$ & 3.13$\times 10^{7}$  & 3$\times 10^{6}$  & 42423  & 505  & 35 &  5.4 $\times 10^{-5}$\%   \tabularnewline
\hline\hline
\end{tabular}
\end{table}

We separate the SM backgrounds according to the flavor of the jet
produced in association with the $t\bar{t}$ pair:
\begin{eqnarray}
t\bar{t}j &:& pp\to t \bar{t} j  \to bW^{+}\bar{b} W^{-}j \to b\bar{b}jjj\ell^+\nu, \\
t\bar{t}b &:& pp\to t \bar{t} b  \to bW^{+}\bar{b} W^{-}b \to b\bar{b}bjj\ell^+\nu.
\label{eq:backg}
\end{eqnarray}
The ``extra" jet in association with the $t\bar{t}$ originates from a light-flavor quark or gluon 
in the first case and from a $b$ or $\bar{b}$ in the second case.   
As shown below, the two backgrounds are suppressed by different kinematic cuts.
In the generation of background events, we demand the transverse momentum ($p_T$) of the 
extra jet to be harder than 10~GeV to avoid soft and collinear divergences from QCD radiation.  
After kinematic cuts, the contributions from 
other SM backgrounds, e.g. $W^+W^-jjj$, are quite small and are 
not included in our analysis.   

For an integrated luminosity of $100~{\rm fb}^{-1}$, the numbers 
of signal and background events at the event generator level are shown in the second 
column of Table~\ref{tab:xsec}.  The top quark decay branching ratio 
${\rm Br}(t\bar{t}\to b\bar{b}\ell^+\nu jj)=2/27$ is included in the numbers.  
We choose six benchmark points for the mass $m_{W^\prime}$.  We set the 
$W^\prime$-$t$-$b$ couplings at $f_{L}=1$ and $f_{R}=0$. 
The rates for other values of $f_{L}$ can be obtained from simple scaling
\be
\sigma = f_{L}^2 \times \sigma({f_{L}=1}).
\ee

\subsection{Selection cuts}

At the analysis level, all the signal and background
events are required to pass the {\it basic} selection cuts listed here:
\bea
&&p_T^j\geq 25\,{\rm GeV}, \qquad 
\left|\eta_{j}\right|\leq 2.5 \nonumber \\
&&p_{T}^{\ell}\geq 25\,{\rm GeV},\qquad
\left|\eta_{\ell}\right|\leq 2.5, \nonumber \\
&&\Delta R_{jj,j\ell,\ell\ell} > 0.4, ~\not{\!{\rm E}}_T > 25~{\rm GeV}
\label{eq:cut1}
\eea
where $p_T$ denotes the transverse momentum, $\not{\!{\rm E}}_T$ is the missing transverse momentum from the invisible neutrino in the final state, and $\Delta R$ is the separation in 
the azimuthal angle ($\phi$)-pseudorapidity ($\eta$) plane between the 
objects $k$ and $l$
\begin{eqnarray} 
\Delta R_{kl}\equiv\sqrt{\left(\eta_{k}-\eta_{l}\right)^{2}
+\left(\phi_{k}-\phi_{l}\right)^{2}}.
\end{eqnarray}
We smear the final state hadronic and leptonic energy according to 
a fairly standard Gaussian-type detector resolution given by
\begin{eqnarray} 
\frac{\delta E}{E}= \frac{\mathcal{A}}{\sqrt{E/{\rm GeV}}}\oplus \mathcal{B},
\end{eqnarray}
where $\mathcal{A}=5 (100)\%$ and $\mathcal{B}=0.55 (5)\%$ for leptons (jets).

\begin{figure*}
\includegraphics[scale=0.35]{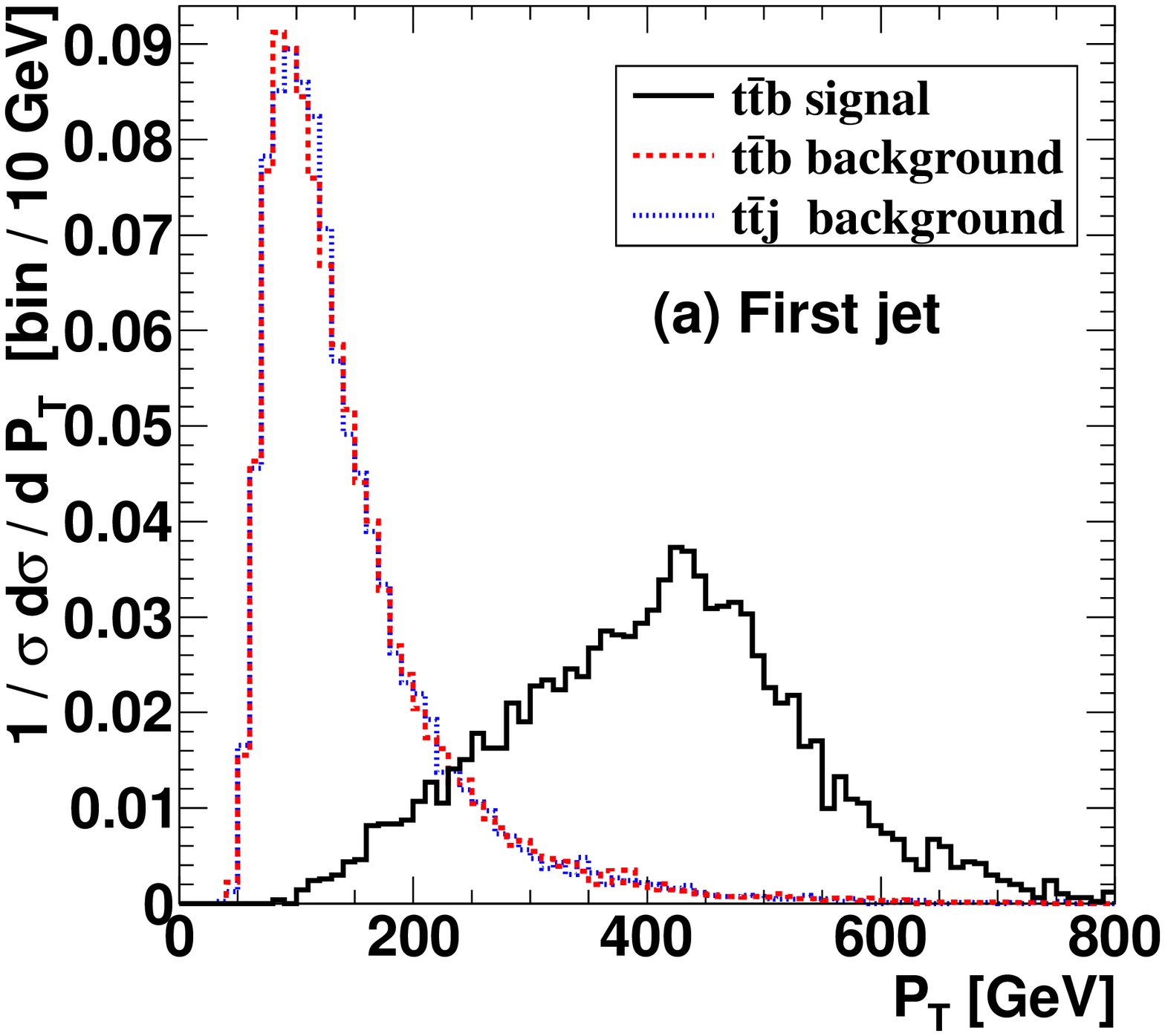}\includegraphics[scale=0.35]{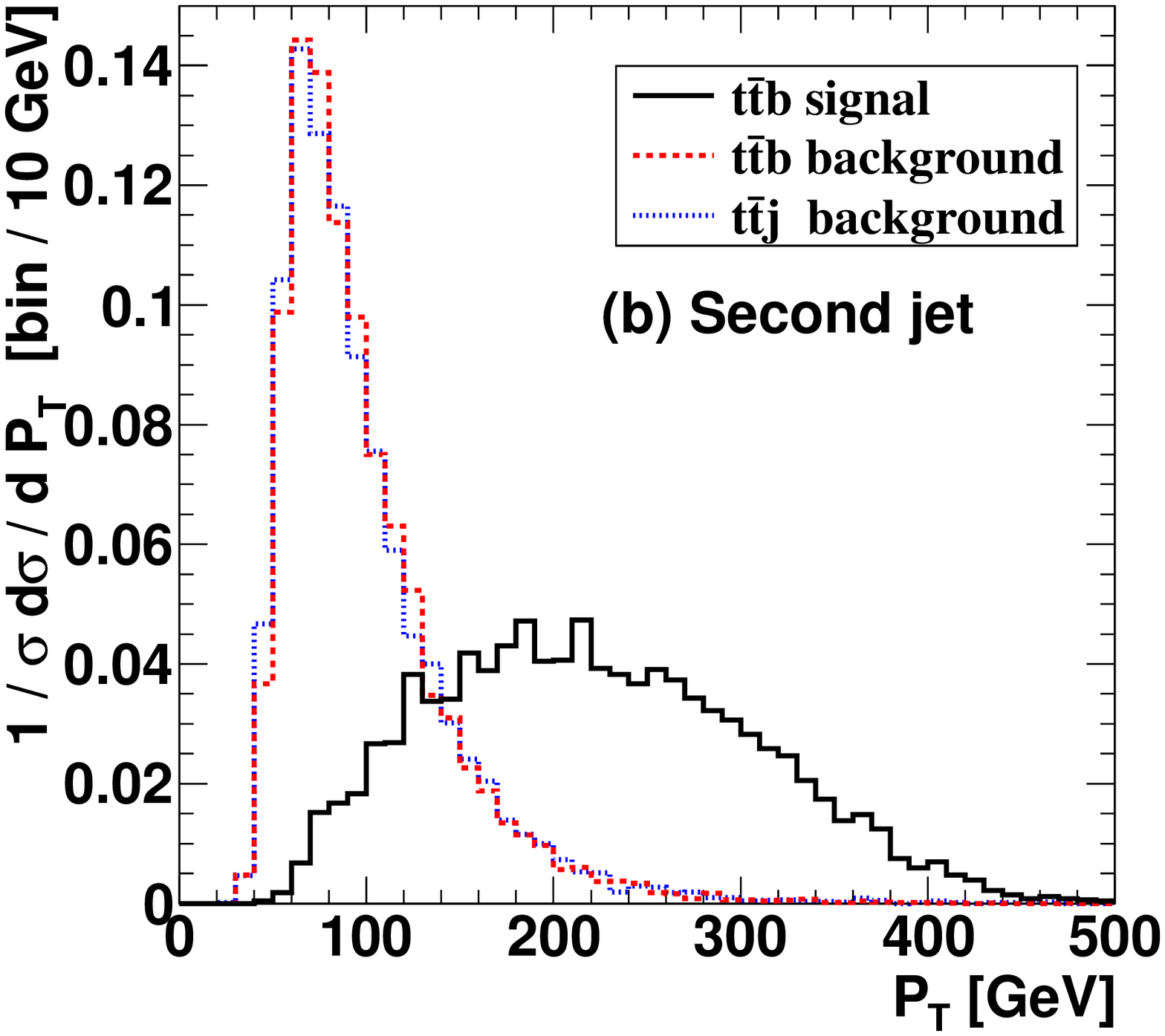}
\includegraphics[scale=0.35]{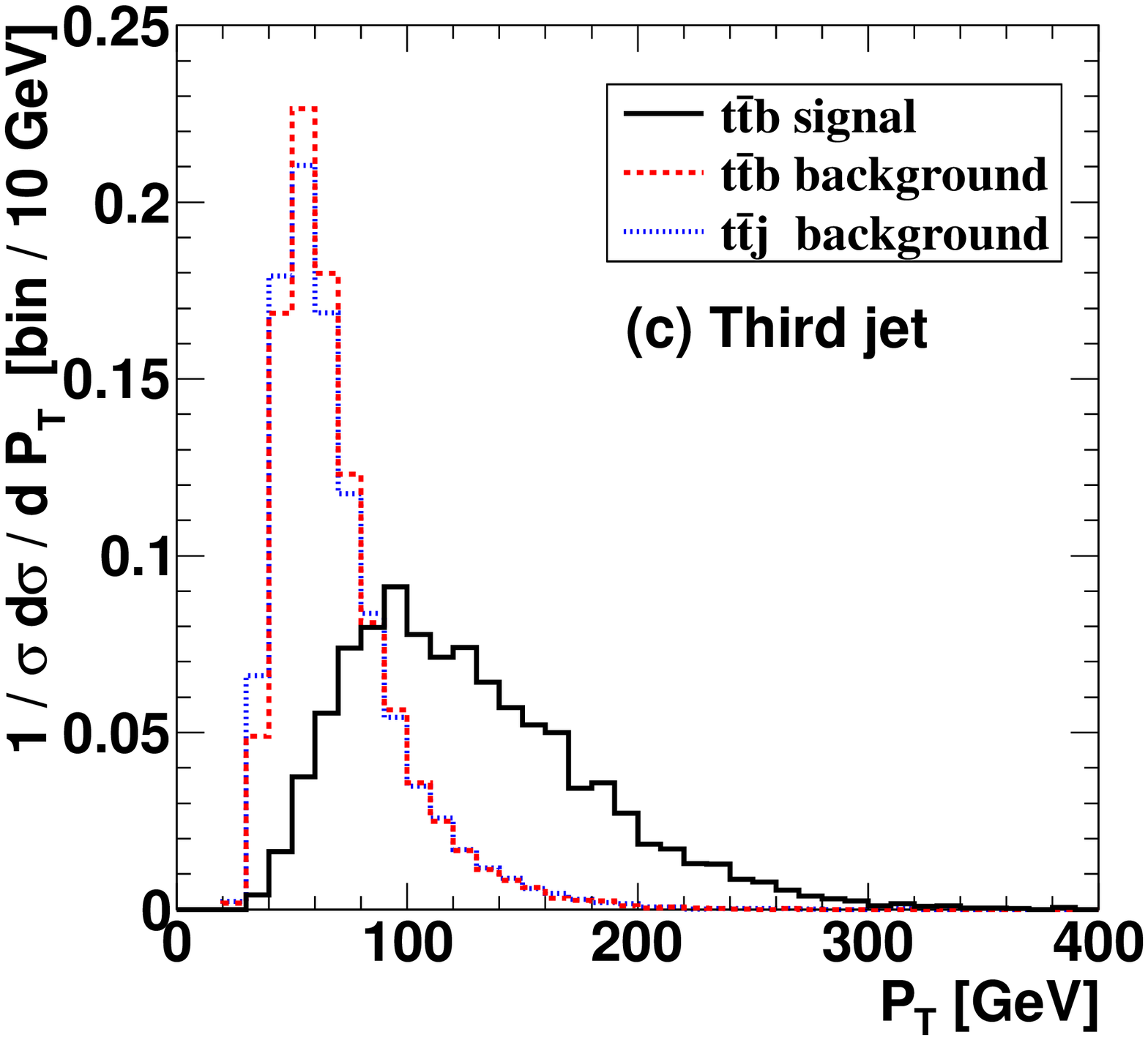}\includegraphics[scale=0.35]{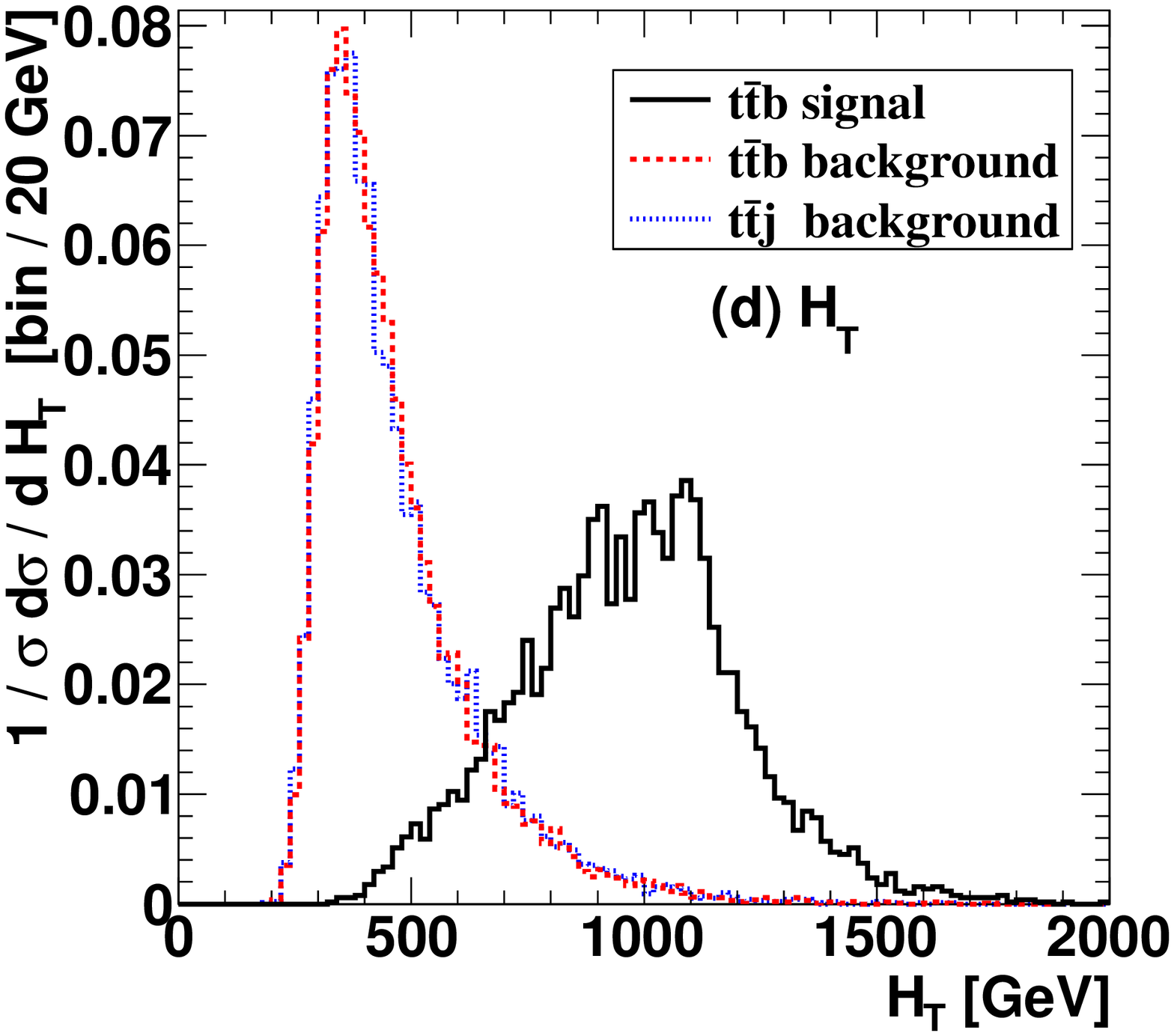}
\caption{Normalized $p_T$ distribution of (a) the leading jet,  (b) the second leading jet, (c)  
the third leading jet, as well as (d) the normalized $H_T$ distribution.  The black-solid curves
represent the signal ($m_{W^\prime}=1~{\rm TeV}$),  the red-dashed curves  
the $t\bar{t}b$ background, while the blue-dotted curves the $t\bar{t}j$ background.
\label{fig:discriminator}}
\end{figure*}

As shown in the third column of Table~\ref{tab:xsec}, roughly 1/3 of the signal events 
pass the basic analysis cuts.
At this stage, the SM backgrounds are dominant over the signal. 
A set of optimized cuts, based on the kinematic differences between the signal and 
backgrounds, is needed to extract the small signal.   

There are five jets in the final state.  Jets from a heavy $W^\prime$ boson 
decay tend to have a harder $p_T$ than jets in the backgrounds.   We order the jets 
by their values of $p_T$.   
Jet charge would also be a possibility for labeling jets, but the charge of jets is not 
well measured experimentally.   Figure~\ref{fig:discriminator}(a) displays the normalized 
$p_T$ distribution of the jet with largest $p_T$ for a 1~TeV $W^\prime$. 
The signal and background curves are normalized by their individual cross sections.   
The signal distribution (black solid curve) peaks around 450~GeV while the backgrounds  
peak around 60-80~GeV.  The leading jet in the signal is mainly the $b$-jet 
from $W^\prime \rightarrow t b$ decay. 
It shares energy with its top quark partner;  therefore its $p_T$ is about $m_{W^\prime}/2$. 
On the other hand, the leading jet in the backgrounds is predominately from top quark 
decay.  Its $p_T$ spectrum peaks around $m_t/3\sim 60~{\rm GeV}$.   These distinct $p_T$ 
spectra motivate a hard cut on the leading jet $p_T$.   

The $p_T$ spectrum of the background is independent of $m_{W^\prime}$ whereas 
the $p_T$ spectrum of the leading jet in the signal is sensitive to $m_{W^\prime}$. 
Absent prior knowledge of $m_{W^\prime}$, a first step in a search might be to 
introduce a mass independent cut to suppress backgrounds, such as to require 
$p_T >  120$~GeV.   This cut could then be increased or decreased to probe for 
a signal, as we expect experimental collaborations will do to search for heavy 
resonances.  

Since the signal strength is much smaller than the background, we 
perform Monte Carlo simulations to find the best cut for each $m_{W^\prime}$.  
This best cut is provided by the simple parameterization 
\be
p_T^{\rm 1st} \geq \left(50.0 + \frac{m_{W^\prime}}{5}\right)~{\rm GeV}, \label{eq:cut2}
\ee
which works well for $400~{\rm GeV} < m_{W^\prime} < 1.0~{\rm TeV}$.  
We think of these $m_{W^\prime}$ dependent cuts as different cut thresholds.
Our $m_{W^\prime}$ dependent cuts are optimized for discovery, and the  
numbers shown in the fourth column (labeled ``optimal'') in Table~\ref{tab:xsec} 
should be viewed as optimized results for each $m_{W^\prime}$. 

Figures~\ref{fig:discriminator}(b) and \ref{fig:discriminator}(c) show the $p_T$ spectra of 
the second and the third leading $p_T$ jets.  Similar to the leading jet, the 2nd and 
3rd leading jets in the signal are harder than those in the backgrounds.   We impose 
kinematic cuts on the 2nd and 3rd jets as follows: 
\bea
p_T^{2nd}&\geq& \left( 20.0 + \frac{m_{W^\prime}}{10} \right)~{\rm GeV}, \nonumber \\
p_T^{3rd}&\geq& \left(20.0 + \frac{m_{W^\prime}}{50}\right)~{\rm GeV}. 
\label{eq:cut3}
\eea

Another useful variable is $H_T$, the scalar sum of the $p_T$'s of all the visible 
particles in the final state, 
\be
H_T= p_T^{\ell^+} + \sum_{j} p_T^j.
\ee
Figure~\ref{fig:discriminator}(d) shows the $H_T$ distributions for the signal and backgrounds. 
Involving a massive $W^\prime$ in the final state, 
the signal distribution peaks above 1~TeV while the background distributions peak near 
the mass threshold of a $t\bar{t}$ pair ($\sim 400~{\rm GeV}$).   This difference enables us to 
impose a hard cut on $H_T$ to further suppress the SM background: 
\be
H_T > \left(m_{W^\prime} - \frac{m_{W^\prime}}{10}\right)~{\rm GeV}, \label{eq:cut4}
\ee

The fourth column of Table~\ref{tab:xsec} shows the number of signal and background events
after the {\it optimized} cuts listed in Eqs.~(\ref{eq:cut2}-\ref{eq:cut4}). 
The $t\bar{t}b$ background is suppressed significantly, but the $t\bar{t}j$ background still 
overwhelms the signal.   However, as we now show, if $b$-tagging can be applied to the extra 
jet (the jet produced in association with the $t\bar{t}$ pair),  the $t\bar{t}j$ background can be 
suppressed efficiently.  This improvement arises because the extra jet in the signal originates 
from the $b$ quark in the $W^\prime$ decay while the extra jet in the $t\bar{t}j$ background is 
from a non-$b$ quark .  
 
\subsection{$\chi^2$-template and extra-jet tagging}

Superficially, the only difference one sees among the final states in 
Eqs.~(\ref{eq:signal}-\ref{eq:backg}) is that the signal and the $t \bar{t} b$ background 
produce final states with 3 $b$ jets, whereas the $t \bar{t} j$ background has only 2 $b$ jets.  
The key to suppressing the $t \bar{t} j$ background is to identify the extra $b$ jet 
in the final state.   To do this, we first exploit the  difference in $p_T$ between the extra jet and 
the other jets, and then we require $b$ tagging to identify it as a $b$ jet.   

The extra jet in the signal comes from the heavy $W^\prime$ decay and tends to have large 
$p_T$.    The extra jet in the SM backgrounds, mainly from QCD radiation,  tends to have 
a much softer $p_T$.   However, a complication is that top quarks in the signal events
are boosted and jets from top quark decay have large $p_T$.  One of the jets 
from top quark decay could play the role of the leading jet.   Our simulation shows
that the extra jet (from heavy $W^\prime$ decay) serves as the leading $p_T$ jet in 
62\% to 84~\% of the cases for $m_{W^\prime}$ ranging from 400~GeV to 1000~GeV.   
In view of small signal rate for a heavy $W^\prime$, a more efficient method is needed 
to identify the extra-jet.  
 
In this study we use a $\chi^2$-template method based on the $W$ boson and top quark masses 
to select the extra jet.  For each event we pick the combination which minimizes the following 
$\chi^2$:
\begin{eqnarray}
	\chi^2 = \frac{(m_W-m_{jj})^2}{\Delta m_W^2} + \frac{(m_t-m_{jl\nu})^2}{\Delta m_t^2} 
	+ \frac{(m_{t}-m_{jjj})^2}{\Delta m_t^2}.
\end{eqnarray}
There is two-fold ambiguity in the reconstruction of  the longitudinal  momentum of the neutrino 
from top quark decay.
Making use of the $W$-boson on-shell condition, $m_{l \nu}^2 = m_W^2$, we can determine 
the longitudinal momentum of the neutrino ($p_{\nu L}$) as 
\begin{eqnarray}
p_{\nu L}  ={1 \over {2 p_{e T}^2}}
 \left( {A\, p_{e L} \pm E_e \sqrt{A^2  - 4\,{p}^{\,2}_{e T}\not{\!\!{\rm E}}_{T}^2}} \right),
\end{eqnarray}
where $A = m_W^2 + 2 \,\vec{p}_{e T} \cdot \,\vec{\not{\!\!{\rm E}}_{T}} $.  If $A^2 - 4  {p}^{2}_{e T} \not{\!\!{\rm E}}_{T}^{2} \geq 0$, 
the value of $p_{\nu L}$ that best yields the known top mass is selected via $m_{l\nu b}^2 = m_t^2$.  
Once detector resolution is taken into account, this ideal situation need not hold.  
In this case, the value of $p_{\nu L}$ is chosen which yields the minimum $\chi^2$.
The reconstruction efficiencies ($\epsilon$) for a $1$~TeV $W^\prime$ compared with 
Monte Carlo truth are found to be:
\bea
\epsilon_{\rm extra} 				  &=& 99.8 \%, \nonumber \\
\epsilon_{\displaystyle t_{\rm lep}}   &=& 98.9  \%, \nonumber \\
\epsilon_{\displaystyle t_{\rm had}}   &=& 92.3  \%. 
\label{eq:chi2-efficiency}
\eea
Such high efficiencies at the parton level arise mainly from the fact that the jets are highly boosted.
Since there are combinatorial ambiguities in the final state, the efficiency for reconstruction of 
a top quark decaying leptonically ($t_{\rm lep}$) is higher than for a top quark decaying 
hadronically ($ t_{\rm had}$).  

Once the extra jet is identified by this kinematic method, one can require it to be a $b$-jet, 
reducing the $t\bar{t}j$ background by about a half, as is shown in the fifth 
column of Table~\ref{tab:xsec}.   To retain as many signal events as possible, we require
only one jet to be $b$-tagged. A tagging efficiency of 60\% is used in our analysis. 
We take into account a mistag rate for a light non-$b$ quark (including the charm quark) 
to mimic a $b$ jet, with mistag efficiency $\epsilon_{j\to b}=0.5\%$. For the 
Monte-Carlo truth events of the signal and backgrounds, we expect that 60\% of the signal and
$t\bar{t}b$ background events pass the $b$-tagging, while $0.5\%$ of the $t\bar{t}j$ background events
pass the $b$-tagging.   Recall that the $t\bar{t}b$ background is suppressed by the 
hard $p_T$ and $H_T$ cuts. The $b$-tagging will further suppress the $t\bar{t}j$ background events with an efficiency $0.5\%$, if one can perfectly identify the extra-jet out of the five jets in the final state.    
However, the extra-jet identification with the $\chi^2$ template method is not perfect.  The jet identified as the extra-jet has three sources: the true extra-jet, $b$-quarks from top (antitop) quark decay, and the light-non-b quark from $W^-$-boson decay.   Multiplying the extra-jet fraction with the corresponding 
jet-tagging efficiency, we show below that one obtains a net jet-tagging efficiency of 1.2~\% for the 
$t\bar{t}j$ background, cf. Eq.~(\ref{eq:b-tag-4}), about twice as large as the case of perfect extra-jet identification (0.5~\%).    

In Table~\ref{tab:rec_eff} we show the 
tagging efficiency $\epsilon_{b{\rm-tag}}$ of the extra jet  after the $\chi^2$-template fit.  It depends on the reconstruction efficiencies for the extra jet:  $\epsilon_{\rm correct}$ denotes the correct fraction 
from the $\chi^2$-fit, $\epsilon_{{\rm wrong-}b}$ is the fraction of $b$ jets from top quark 
decay that fake the extra jet,  while $\epsilon_{{\rm wrong-light}}$ is the fraction of 
light jets from top quark decay that fake the extra jet.  As an example, consider the $b$-tagging 
efficiency in the signal process with 1 TeV $W^\prime$ mass.  Since there are five jets in the final 
state, it is possible that after event reconstruction the extra jet is a $b$-jet from the top quark or 
anti-top quark decay (which we label ``${\rm wrong-}b$"), or a light-flavor jet from hadronic top quark decay (which we label ``${\rm wrong-light}$"), or a $b$-jet from the 
$W^\prime$ decay (which we label ``${\rm correct}$").  Note that  the $b$-tagging is applied to the 
extra jet (which we call ``extra-j-tagging", to avoid confusion with the original $b$-tagging), but not
to the truth $b$-jet from the $W^\prime$ decay. Taking the reconstruction efficiencies into account, 
we evaluate the net $b$-tagging efficiency of the extra jet $\epsilon_{{\rm extra-j-tag}}$ as 
\be
\epsilon_{{\rm extra-j-tag}} = (\epsilon_{\rm correct}+ \epsilon_{{\rm wrong-}b}) \times 0.6 
+\epsilon_{\rm wrong-light}\times 0.005 \,,
\label{eq:b-tag-1}
\ee
for the signal process.  A similar analysis gives us the same formula for the $t\bar{t}b$ 
background.    For a 1~TeV $W^\prime$ with hard cuts,
we find the extra jet is a $b$-jet  with 99.9\% probability, and a light jet with 
0.4\% probability. Therefore, the $b$-tagging efficiency for the signal  is 
\be
0.999\times 0.6 + 0.004\times 0.005 = 0.60.  
\label{eq:b-tag-3}
\ee
For the $t\bar{t}j$ background, the correct jet is a light-flavor jet from the real radiation 
associated with top pair production.  The formula changes to
\be
\epsilon_{{\rm extra-j-tag}} = (\epsilon_{\rm correct}+\epsilon_{\rm wrong-light} ) \times 0.005 
+\epsilon_{{\rm wrong-}b}\times 0.6 
\label{eq:b-tag-2}
\ee
for the $t\bar{t}j$ background. 
For the $t\bar{t}j$ background, in the case of a 1~TeV $W^\prime$ with hard cuts, 
we find extra jet is a light jet with 98.8\% probability, and a $b$-jet with 
1.18\% probability.   Therefore, the $b$-tagging efficiency for the $t\bar{t}j$ 
background is 
\be
0.988\times 0.005 + 0.012\times 0.6 = 0.012.  
\label{eq:b-tag-4}
\ee  
We use Eqs.~(\ref{eq:b-tag-1}-\ref{eq:b-tag-4}) to explain the numbers in the 
$b$-tagging column in Table~\ref{tab:rec_eff}.   Table~\ref{tab:rec_eff} shows 
that about 60\% of the signal and $t\bar{t}b$ background events pass $b$-tagging 
even with the imperfect reconstruction of the extra jet.  However, the extra-jet tagging 
efficiency for  the $t\bar{t}j$ background is always larger than the $b$-tagging 
efficiency of $0.005$ because it always is possible to mistag a $b$-jet when 
the extra-jet tagging is done.

\begin{table}
\caption{The efficiency for extra jet reconstruction with the $\chi^2$-template method. 
The net $b$-tagging efficiencies ($\epsilon_{{\rm extra-j-tag}}$), calculated with Eqs.~\ref{eq:b-tag-1}
and~\ref{eq:b-tag-2}, are shown in the last column. 
\label{tab:rec_eff}}
\begin{tabular}{ccccccc}
\hline\hline
$m_{W^\prime}$ &  $\epsilon_{\rm correct}$  & $\epsilon_{{\rm wrong-}b}$ & $\epsilon_{\rm wrong-light}$  & $\epsilon_{{\rm extra-j-tag}}$    \tabularnewline 
\hline\hline
400          &   98.15 \%  & 1.63 \% & 0.22 \% & 59.9 \% \tabularnewline
 $t\bar{t}b$ &  98.34 \%   & 1.5 \%  & 0.15 \% &  59.9 \% \tabularnewline
 $t\bar{t}j$ &  96.67 \%   & 2.96 \% & 0.37 \% &  2.26 \%   \tabularnewline
\hline
500          &   98.53 \%  & 1.35 \% & 0.12 \% & 59.9 \% \tabularnewline
 $t\bar{t}b$ &  98.34 \%   & 1.5 \%  & 0.16 \% &  59.9 \% \tabularnewline
 $t\bar{t}j$ &  96.7 \%   & 2.92 \% & 0.37 \% &  2.23 \%   \tabularnewline
\hline
600          &   99.32 \%  & 0.59 \% & 0.08 \% & 59.9 \% \tabularnewline
 $t\bar{t}b$ &  98.34 \%   & 1.48 \%  & 0.15 \% &  59.9 \% \tabularnewline
 $t\bar{t}j$ &  96.75 \%   & 2.88 \% & 0.36 \% &  2.22 \%   \tabularnewline
\hline
700          &   99.4 \%  & 0.51 \% & 0.09 \% & 59.9 \% \tabularnewline
 $t\bar{t}b$ &  98.6 \%   & 1.27 \%  & 0.12 \% &  59.9 \% \tabularnewline
 $t\bar{t}j$ &  97.15 \%   & 2.55 \% & 0.29 \% &  2.02 \%   \tabularnewline
\hline
800          &   99.66 \%  & 0.31 \% & 0.03 \% & 60 \% \tabularnewline
 $t\bar{t}b$ &  98.93 \%   & 1.0 \%  & 0.07 \% & 60 \% \tabularnewline
 $t\bar{t}j$ &  97.6 \%   & 2.17 \% & 0.23 \% &  1.79 \%   \tabularnewline
\hline
900          &   99.87 \%  & 0.12 \% & 0.01\% & 60 \% \tabularnewline
 $t\bar{t}b$ &  99.24 \%   & 0.69 \%  & 0.06 \% &  60 \% \tabularnewline
 $t\bar{t}j$ &  98.17 \%   & 1.63 \% & 0.2 \% &  1.46 \%   \tabularnewline
\hline
1000          &   99.79 \%  & 0.14 \% & 0.06 \% & 60 \% \tabularnewline
 $t\bar{t}b$ &  99.5 \%   & 0.43 \%  & 0.07 \% &  60 \% \tabularnewline
 $t\bar{t}j$ &  98.65 \%   & 1.18 \% & 0.17 \% &  1.2 \%   \tabularnewline
\hline\hline
\end{tabular}
\end{table}

\subsection{Mass window $\Delta M$ cut}

\begin{figure}
	\includegraphics[scale=0.35]{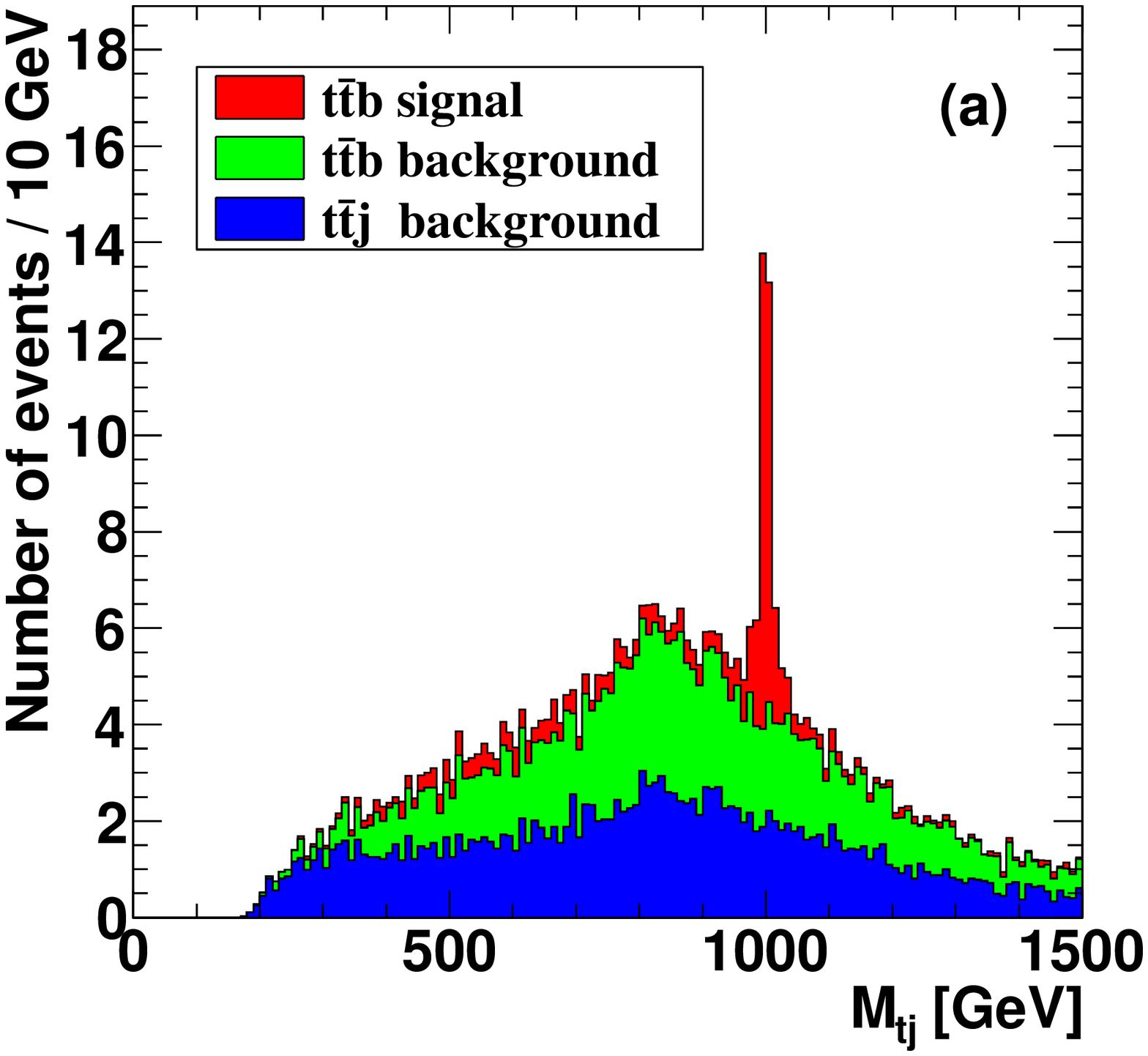}
	\includegraphics[scale=0.35]{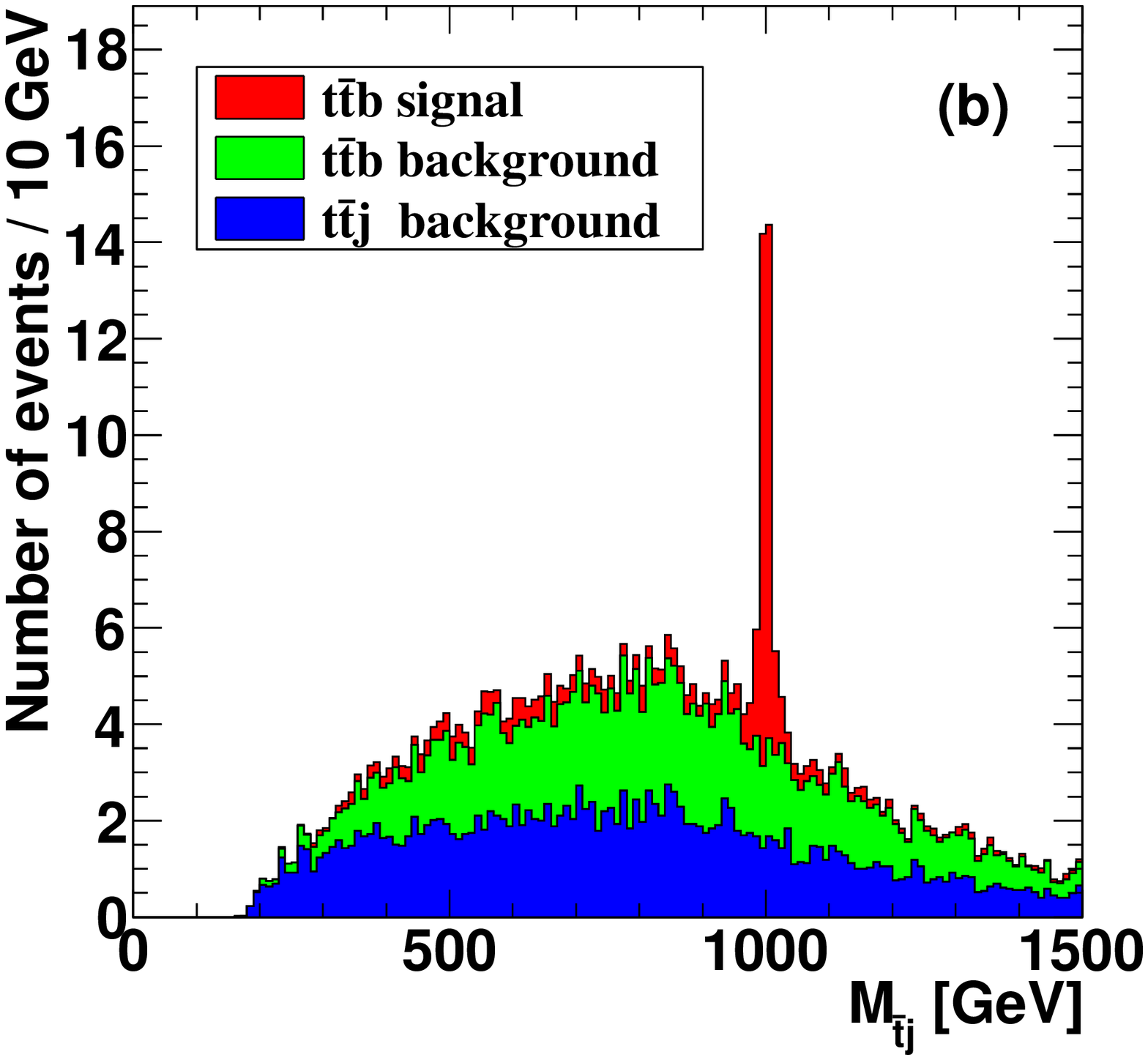}
	\caption{(a) Reconstructed invariant mass distribution of the leptonic decaying top and 
	extra jet; (b) reconstructed hadronic decaying top and extra jet invariant mass distribution. \label{fig:mass}}
\end{figure}

After full event reconstruction, one can compute the $W^\prime$ mass formed from 
the extra jet and the reconstructed $t$- or $\bar{t}$-quark. 
Since our signal events consist of both $tW^{\prime-}$ and $\bar{t}W^{\prime+}$,
one half of the signal events exhibit a peak in the invariant mass spectrum of 
the extra jet and $t$ quark (denoted as $m_{tj}$) while the other half have a peak 
in the invariant mass of the extra jet and $\bar{t}$ quark (denoted as $m_{\bar{t}j}$). 
Figure~\ref{fig:mass} shows the reconstructed $m_{tj}$ and $m_{\bar{t}j}$ distributions
for the signal (red), $t\bar{t}j$ (blue) and $t\bar{t}b$ (green) backgrounds.
The signal distribution shows a sharp peak at the input value of $m_{W^\prime}$.
The pin shape reflects the narrow width of the top-philic $W^\prime$ boson, 
e.g. the $W^\prime$ width is about 8~GeV for a 1~TeV $W^\prime$.  
The long tail into the small mass region comes from the one-half wrong combination.   
The peaks of the background distributions around 800~GeV are caused by 
the combination of hard kinematic cuts and jet identification (with combinatorial factors 
included).  

Once $m_{W^\prime}$ is known, we can impose cuts on $m_{tj}$ or $m_{\bar{t}j}$
to further suppress backgrounds. 
We first demand large invariant masses for both $tj$ and $\bar{t}j$,
\be
m_{tj} > 250 +  \frac{m_{W^\prime}}{4},\qquad 
m_{\bar{t}j} > 250 +  \frac{m_{W^\prime}}{4},
\ee
and that one of the following two mass window cuts be satisfied,
\be
\left|m_{tj} - m_{W^\prime}\right| < \frac{m_{W^\prime}}{10},
~~{\rm or}~~\left|m_{\bar{t}j} - m_{W^\prime}\right| < \frac{m_{W^\prime}}{10}. 
\ee
The mass window suppress both SM backgrounds by a factor of 10 while it keeps
most of the signal.

\subsection{Discovery potential}

\begin{figure}
\includegraphics[scale=0.35]{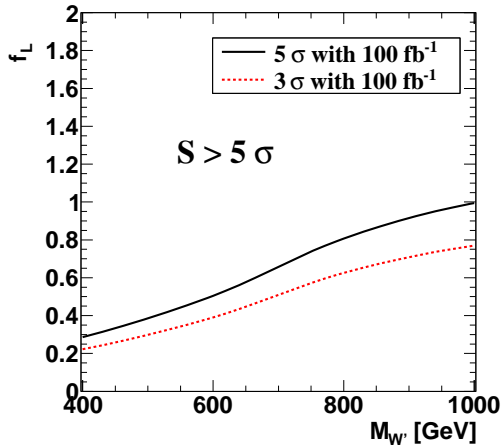}
\caption{The discovery potential for the top-philic $W^\prime$ 14~TeV with an integrated 
luminosity of $100~{\rm fb}^{-1}$. }
\label{fig:discovery}
\end{figure}

The SM backgrounds are suppressed efficiently such that less than 1 background event survives after cuts with an integrated luminosity of $100~{\rm fb}^{-1}$. 
For a 1~TeV $W^\prime$ with the same coupling strength as the SM $W$-$t$-$b$ interaction, 
we obtain a $5$ standard deviations ($\sigma$) 
statistical significance, 
defined as $S/\sqrt{B}$ where $S$ and $B$ denotes the number of signal and background events, respectively. 
For a lighter $W^\prime$, the significance is larger for fixed coupling strength.
The $3~\sigma$ and $5~\sigma$ discovery curves are plotted in Fig.~\ref{fig:discovery}. The 
region above the $5~\sigma$ curve is good for discovery. 

\section {$W^\prime$-$t$-$b$ coupling and $t$-polarization}
After the discovery of this $W'$ boson, one would like to know its mass, spin, and 
couplings.  The invariant mass or transverse momentum distributions of its decay 
products can be used to determine its mass.
Angular distributions of its decay products can be investigated to
confirm its spin and the chiral structure of the $W'$ couplings to SM 
fermions. 
\begin{figure}[t]
	\includegraphics[scale=0.35]{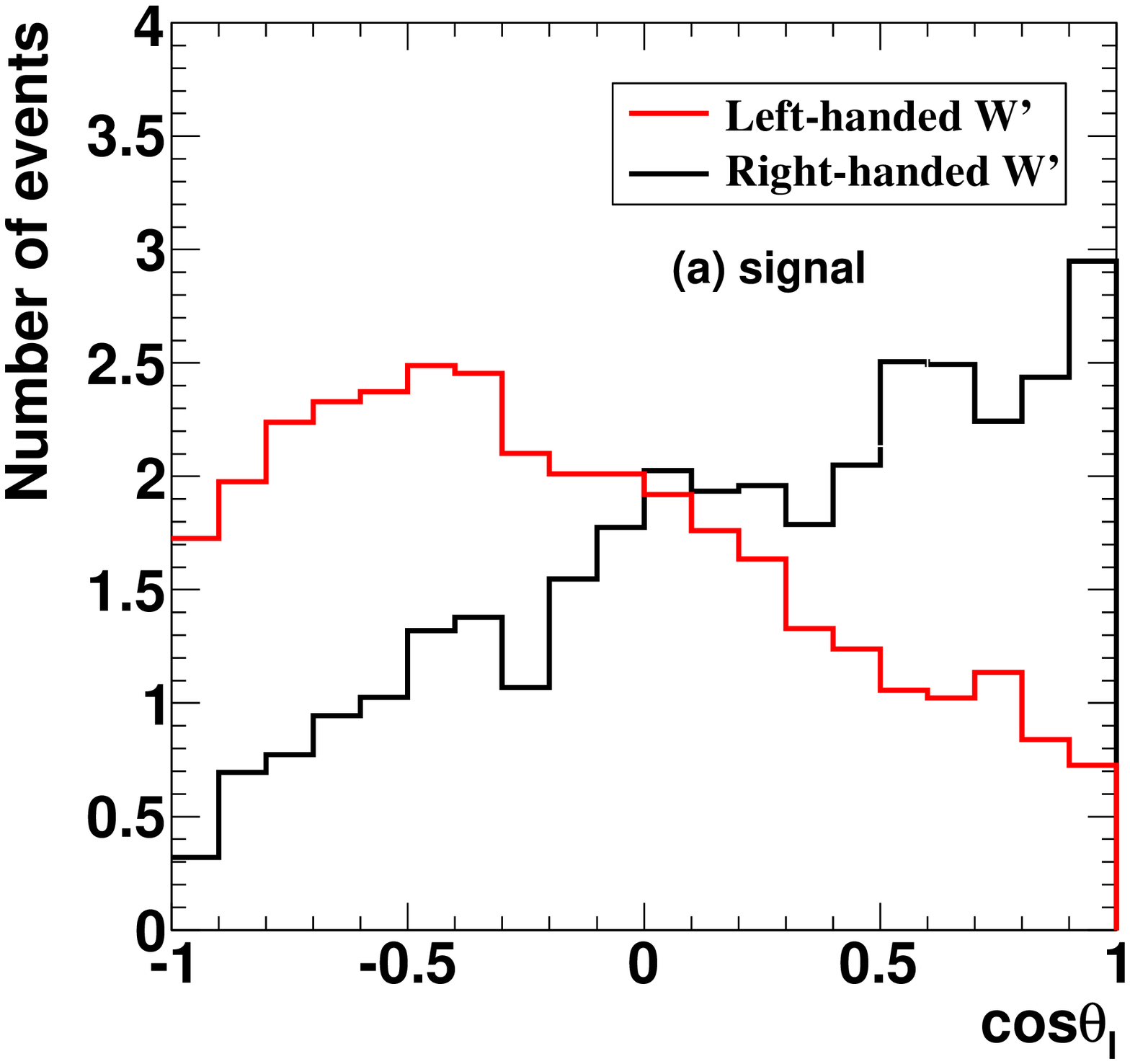}
	\includegraphics[scale=0.35]{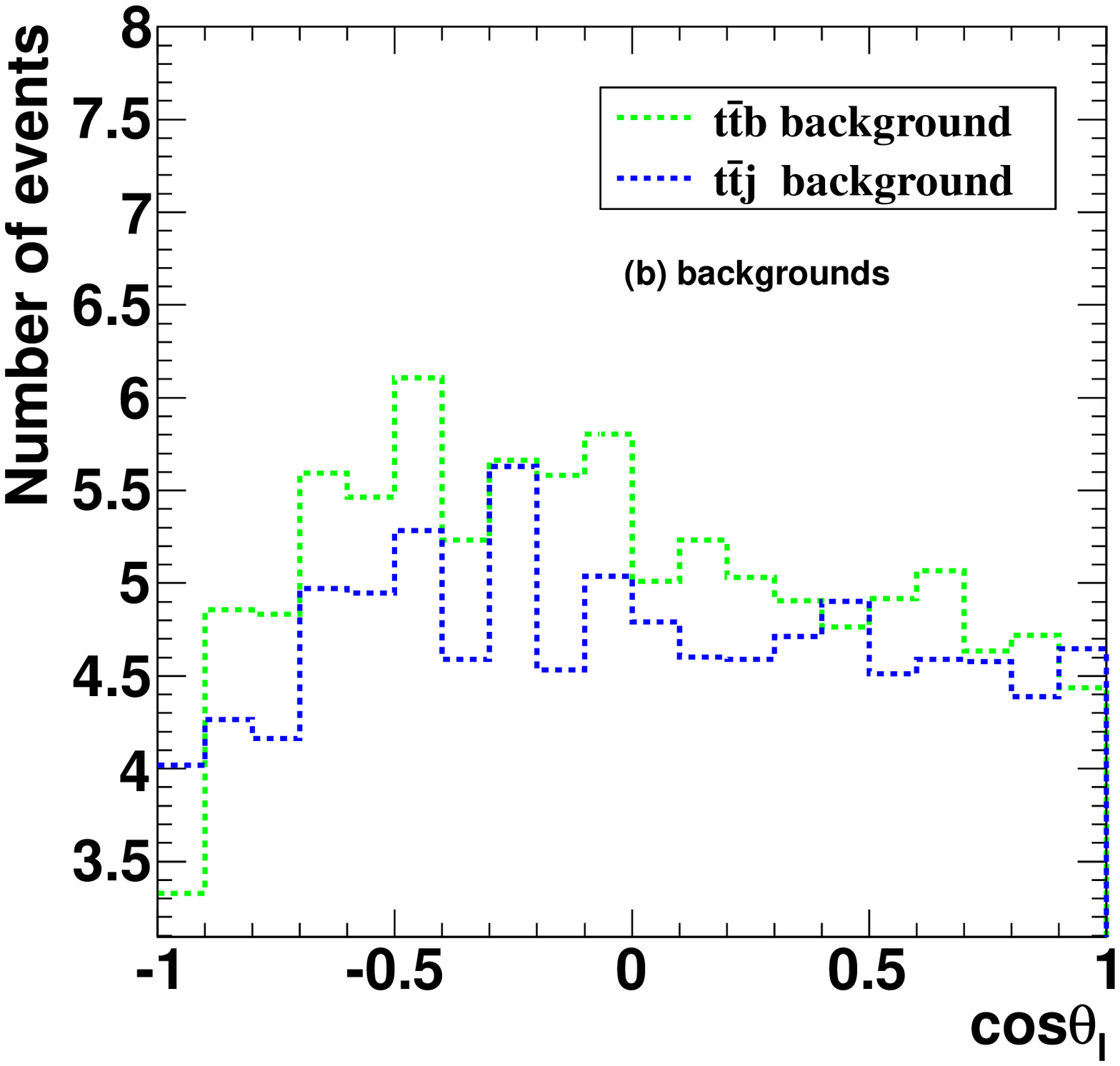}
	\caption{The angular distributions of the final lepton $\cos\theta_l$ for the left-handed and 
	the right-handed $W'$.}
	\label{fig:pol}
\end{figure}
The chirality of the $W'$ coupling to SM 
fermions is best measured from the polarization of the top quark~\cite{Gopalakrishna:2010xm,Berger:2011hn}.
Among the top quark decay products, the charged lepton from $t \to b l\nu$
is the best analyzer of the top quark spin.
For a left-handed top quark, the charged lepton moves 
preferentially against the direction of motion of the top quark, 
while for a right-handed top quark the charged lepton moves along 
the direction of motion of the top quark.  The angular correlation of the lepton is
$\frac{1}{2}(1\pm \cos\theta_l)$, with the $(+)$ choice for right-handed and $(-)$ 
for left-handed top quarks, where
$\theta_l$ is the angle of the lepton 
in the rest frame of top quark relative to the top quark direction of motion 
in the center-of-mass (cm) frame of the incoming partons.   
In Fig.~\ref{fig:pol} we plot the $\cos\theta_l$ distribution for $f_L=1,~f_R=0$ and 
$f_L=0, f_R=1$ couplings. The curves clearly show the main characteristic features of 
the $\frac{1}{2}(1\pm \cos\theta_l)$ 
behaviors for purely right- and left-handed polarized top quarks from $W'$ decay,  
even after kinematic cuts are imposed. 
We note that due to the $p_T$ and $\Delta R$ cuts, the distributions are distorted 
and drop significantly in the region $\cos\theta_l \sim -1$ for $f_L = 1$ and $f_R = 0$, 
and $\cos\theta_l \sim 1$ for $f_R =1$ and $f_L =0$.
We expect a flat angular distribution for the SM background because the top quark 
and anti-top quark are not polarized.  
Therefore, the angular distributions of the lepton can be used to discriminate
top-philic $W'$ models in which the  chirality of the $W'$ coupling 
to SM fermions differs.

\section{Conclusion}

In this paper we examine the LHC phenomenology of a top-philic $W^\prime$ model.   
In the model the $W^\prime$ boson is produced in association with a top-quark and it 
decays into a top quark and bottom quark pair, yielding a collider signature of 
$t\bar{t}$ plus one $b$-jet.   We exploit the different kinematic features of the signal and 
backgrounds to suppress the large standard model backgrounds from $t\bar{t}j$ and 
$t\bar{t}b$ production.  Examining the distributions of the signal and backgrounds, we 
find that hard $p_T$ cuts and cuts on $H_T$ can suppress the $t\bar{t}b$ 
background.   After full event reconstruction, we show that tagging the extra $b$-jet can 
further suppress the $t\bar{t}j$ background.   We show that discovery of a top-philic 
$W^\prime$ with SM-like coupling strengths is promising at 14 TeV with 
${\cal{L}} = 100\,\,fb^{-1}$.  A resonance peak in the top quark and $b$-jet invariant 
mass distribution is a distinct signature of $W^\prime$ discovery.
Top quark polarization can be used to measure the chiral structure of the 
$W^\prime$-$t$-$b$ coupling.   Top quark pair and hard $b$-jet final states are  
worth examining even in a model-independent way.   This final state is a new 
unexploited channel at the LHC.

\begin{acknowledgments}
The work by E. L. B. and Q.H.C. is supported in part by the U.S.
DOE under Grants No.~DE-AC02-06CH11357. Q.H.C. is also supported in part 
by the Argonne National Laboratory and University
of Chicago Joint Theory Institute Grant 03921-07-137. 
The work by J.H.Y. and C.P.Y. is supported in part by the U.S. National Science Foundation 
under Grand No. PHY-0855561.
E.L.B thanks the Kavli Institute for Theoretical Physics (KITP), Santa Barbara, for hospitality while this research was being completed.  Research at KITP is supported in part by the National Science Foundation under Grant No. NSF PHY05-51164.   J.H.Y. thanks Reinhard Schwienhorst for discussions of $b$-tagging and Argonne National Laboratory for hospitality during several visits while part of work was being done.
\end{acknowledgments}


\begin{thebibliography}{199}

\bibitem{Khachatryan:2010fa}
  V.~Khachatryan {\it et al.}  [CMS Collaboration],
  Phys.\ Lett.\  B {\bf 698}, 21 (2011)
  [arXiv:1012.5945 [hep-ex]].

\bibitem{Aad:2011fe}
  G.~Aad {\it et al.}  [ATLAS Collaboration],
  arXiv:1103.1391 [hep-ex].

\bibitem{Aaltonen:2010jj}
  T.~Aaltonen {\it et al.} [ CDF Collaboration ],
  Phys.\ Rev.\  {\bf D83}, 031102 (2011).
  [arXiv:1012.5145 [hep-ex]].


\bibitem{Aaltonen:2009qu}
  T.~Aaltonen {\it et al.} [ CDF Collaboration ],
  Phys.\ Rev.\ Lett.\  {\bf 103}, 041801 (2009).
  [arXiv:0902.3276 [hep-ex]].


\bibitem{Abazov:2010dj}
  V.~M.~Abazov {\it et al.}  [D0 Collaboration],
  arXiv:1011.6278 [hep-ex].

\bibitem{Aaltonen:2011kc}
  T.~Aaltonen {\it et al.}  [CDF Collaboration],
  Phys.\ Rev.\  D {\bf 83}, 112003 (2011)
  [arXiv:1101.0034 [hep-ex]].

\bibitem{Barger:2010mw}
  V.~Barger, W.~-Y.~Keung, C.~-T.~Yu,
  Phys.\ Rev.\  {\bf D81}, 113009 (2010).
  [arXiv:1002.1048 [hep-ph]].
  
\bibitem{Cao:2010zb}
  Q.~H.~Cao, D.~McKeen, J.~L.~Rosner, G.~Shaughnessy and C.~E.~M.~Wagner,
  Phys.\ Rev.\  D {\bf 81}, 114004 (2010)
  [arXiv:1003.3461 [hep-ph]].

\bibitem{Barger:2011ih}
  V.~Barger, W.~-Y.~Keung, C.~-T.~Yu,
  Phys.\ Lett.\  {\bf B698}, 243-250 (2011).
  [arXiv:1102.0279 [hep-ph]]



\bibitem{Gresham:2011dg}
  M.~I.~Gresham, I.~W.~Kim and K.~M.~Zurek,
  arXiv:1102.0018 [hep-ph].
  
  
\bibitem{Malkawi:1996fs}
  E.~Malkawi, T.~M.~P.~Tait and C.~P.~Yuan,
  Phys.\ Lett.\  B {\bf 385}, 304 (1996)
  [arXiv:hep-ph/9603349].

\bibitem{Li:1981nk}
  X.~Li and E.~Ma,
  Phys.\ Rev.\ Lett.\  {\bf 47}, 1788 (1981).

\bibitem{He:2002ha}
  X.~G.~He and G.~Valencia,
  Phys.\ Rev.\  D {\bf 66}, 013004 (2002)
  [Erratum-ibid.\  D {\bf 66}, 079901 (2002)]
  [arXiv:hep-ph/0203036].

\bibitem{Hsieh:2010zr}
  K.~Hsieh, K.~Schmitz, J.~H.~Yu and C.~P.~Yuan,
  Phys.\ Rev.\  D {\bf 82}, 035011 (2010)
  [arXiv:1003.3482 [hep-ph]].

\bibitem{Aaltonen:2009iz}
  T.~Aaltonen {\it et al.}  [CDF Collaboration],
  Phys.\ Rev.\ Lett.\  {\bf 102}, 222003 (2009)
  [arXiv:0903.2850 [hep-ex]].


\bibitem{Berger:2011hn}
  E.~L.~Berger, Q.~H.~Cao, C.~R.~Chen and H.~Zhang,
  Phys.\ Rev.\  D {\bf 83}, 114026 (2011)
  [arXiv:1103.3274 [hep-ph]].

\bibitem{Gopalakrishna:2010xm}
  S.~Gopalakrishna, T.~Han, I.~Lewis, Z.~g.~Si and Y.~F.~Zhou,
  Phys.\ Rev.\  D {\bf 82}, 115020 (2010)
  [arXiv:1008.3508 [hep-ph]].

\bibitem{Alwall:2011uj}
  J.~Alwall, M.~Herquet, F.~Maltoni, O.~Mattelaer and T.~Stelzer,
  arXiv:1106.0522 [hep-ph].
  
\end{thebibliography}
\end{document}